\documentclass{article}
\usepackage{spconf,amsmath,graphicx}
\usepackage{url}
\usepackage{multirow}

\usepackage{makecell}
\usepackage{multicol}
\usepackage{bm}
\usepackage{hyperref}
\usepackage{footnote}
\usepackage{tablefootnote}
\usepackage{subcaption}
\usepackage{amsmath}
\usepackage{amssymb}
\usepackage[dvipsnames]{xcolor}

\title{Enhancing Synthetic Training Data for Speech Commands: \\From ASR-Based Filtering to Domain Adaptation in SSL Latent Space}

\name{Sebastião Quintas$^1$, Isabelle Ferrané$^1$, Thomas Pellegrini$^1$}
\address{$^1$IRIT, Université de Toulouse, CNRS, Toulouse INP, UT3, Toulouse, France\\}
\begin{document}
\ninept
\maketitle
\begin{abstract}
The use of synthetic speech as data augmentation is gaining increasing popularity in fields such as automatic speech recognition and speech classification tasks. Despite novel text-to-speech systems with voice cloning capabilities, that allow the usage of a larger amount of voices based on short audio segments, it is known that these systems tend to hallucinate and oftentimes produce bad data that will most likely have a negative impact on the downstream task. In the present work, we conduct a set of experiments around zero-shot learning with synthetic speech data for the specific task of speech commands classification. Our results on the Google Speech Commands dataset show that a simple ASR-based filtering method can have a big impact in the quality of the generated data, translating to a better  performance. Furthermore, despite the good quality of the generated speech data, we also show that synthetic and real speech can still be easily distinguishable when using self-supervised (WavLM) features, an aspect further explored with a CycleGAN to bridge the gap between the two types of speech material.
\end{abstract}
\section{Introduction}
\label{sec:intro}

Recent works show that text-to-speech (TTS) systems, can effectively be used as a successful data augmentation scheme for automatic speech recognition (ASR) \cite{tts_asr1,tts_asr3,tts_asr4,tts_asr5,tts_asr6}, specially in low resource scenarios \cite{tts_asr_lowres2,tts_asr_lowres3}. Moreover, it is worth nothing that there may be limits to the benefits of using synthetic data, as it has been shown that bispectral analysis can still differentiate with a good degree of confidence between state-of-the-art TTS systems and real human speech \cite{real_vs_tts}, showing a mismatch between the two types of audio. A similar case can be found when using self-supervised (SSL) features in anti-spoofing detection \cite{wavlm_spoofing1}, showing a good discriminating ability between the two types of speech. As a consequence, there is still a noticeable difference between systems trained on synthetic speech and those trained on real data \cite{gap_tts_asr}, leaving an interesting research gap to explore. 

Current TTS systems allow the possibility of multi-speaker synthesis by simply using short reference audio files \cite{xtts_v2_github}. On one hand, the usage of these voice cloning technologies \cite{alexa_kws}, allows the employment of significantly larger corpora and speech resources as generative material. On the other hand, it is known that autoregressive speech synthesis can hallucinate, an aspect that can oftentimes make the generated data low-quality or even unusable \cite{tts_mis2}. The lack of control on these hallucinations can be detrimental to the quality of the generated data, as well as to the downstream task that makes use of it. 

Despite the extensive literature on TTS as a data augmentation technique useful for ASR, there is only a couple of articles on the topic of data augmentation, and more specifically synthetic audio for speech commands classification (SCC) and keyword spotting (KWS). The work of \cite{kws_rnn} fine-tunes a RNN-T KWS system on new keywords obtained from a multi-speaker TTS. The results show considerable improvements on those keywords, however the performance improvements are dependent on speaker diversity and amount of data. In \cite{kws_ssl}, a self-supervised learning (SSL) approach for KWS was used. Here, the system was trained on a large general speech corpora of untranscribed audio, containing noise and reverb as data augmentation. The results show that with this pre-training, the quantity of real data used can be greatly reduced, while still achieving a good performance. Similarly, the work of \cite{kws_limited} also proposes a SSL learning structure for downstream KWS fine-tuning. Here, a large amount of keywords are obtained from well-known public corpora to train a keyword embedding extractor. The results suggest a good adaptability and performance gains when applied to new classes as well as an easy extrapolation to different languages other than English. Similarly, the work of \cite{google_emb} also tested a SSL hypothesis for KWS, however this time by pairing synthetic data with normal data as an augmentation. Here, a pre-trained embedding model was fine tuned on the downstream task of KWS, using different mixes of real and synthetic data, showcasing high accuracy.

In the present work, we intend to explore different topics around SCC systems fully trained on synthetic data, such as: \textbf{(i)} Propose an ASR filtering method that is able to generate hallucination-free synthetic data. \textbf{(ii)} Show that despite the clean synthetic data, there is still a performance gap, and that synthetic speech is still easy to differentiate from real speech using SSL features. \textbf{(iii)} Propose a method to improve the quality of synthetic speech representations in higher dimensional spaces, using the same SSL features, that can further improve SCC performance.


\section{Synthetic spoken commands generation}
\label{synthetic}


We propose a framework to generate synthetic speech data using the open-source implementation of the XTTS v2 speech synthesizer and its zero-shot voice cloning capabilities~\cite{xtts_v2_github}. In order to ensure a large variety of cloned speakers, our method randomly samples, without reposition, spoken utterances gathered from the Common Voice (CV) open source and multilingual initiative~\cite{commonvoice}. CV features a collection of read sentences in a variety of languages. We gathered together a total of 4.36 million utterances spoken by 174 thousand distinct speakers, from the CV datasets available in English, Spanish, French, Italian, German, Mandarin and Japanese. It allowed to use a voice only once for each synthesized audio snippet. 
Preliminary experiments showed us that using multilingual speech data is on-par with using speech data from the target language only.


\subsection{ASR-based filtering in the TTS loop}
\label{syn_gen}

\begin{figure}[!t]
\centering
\includegraphics[width=3.3in]{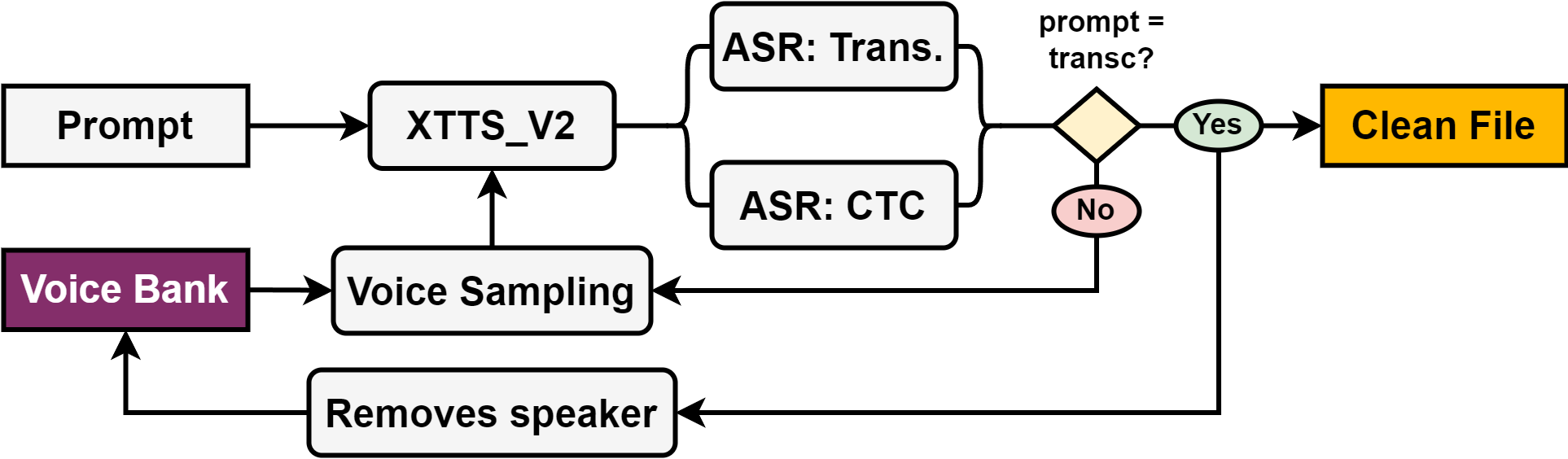}    
\caption{Proposed TTS generation loop with ASR filtering.}
\vspace{-0.7cm}
\label{asr_filtering}
\end{figure}

Given the end-to-end nature of XTTS v2, that can oftentimes produce hallucinations and artefacts, we introduce an ASR filtering scheme, illustrated in Fig. \ref{asr_filtering}. We keep audio snippets for which the two transcriptions obtained with two distinct large vocabulary ASR systems exactly match the word of interest. The two ASR systems are a fast-conformer transducer-based and a Jasper CTC-based models, both trained for English and available off-the-shelf within the NeMO toolkit \cite{nemo}. 
Using two ASR systems improved the filtering, compared to using a single one.

\subsection{Google Speech Commands (GSC)}
\label{speech_commands}


We perform our experiments based on the Google Speech Commands (v2) Dataset \cite{speech_commands}, widely used to benchmark KWS systems. The corpus contains 105,000 utterances, one-second long each, belonging to one of 35 classes. Among these classes lie the ten speech commands (\textit{Yes, No, Up, Down, Left, Right, On, Off, Stop, Go}), followed by numbers from zero to nine and other arbitrary words that covered different phonemes (\textit{Backward, Bed, Bird, Cat, Dog, Forward, Follow, Happy, House, Learn, Marvin, Sheila, Tree, Visual, Wow}). In the present work, we focus on the subset of labels containing the original ten speech commands, categorizing the remaining words under an \textit{unknown} class. While GSC also contains silence audio files, the proposed way to generate these silences is random extraction from a small set of larger files, therefore, it is likely that the same silence slice will end up in both train and validation or test sets, biasing the performance and not promoting a static train, validation and test sets. We believe that this aspect is oftentimes ignored when benchmarking with this dataset. Therefore, we opted to not use the silence class in our experiments.  

\subsection{Two synthetic GSC versions: \textit{Synth} and \textit{Synth (F)}}

We used XTTS v2 with voice cloning on CV, as explained above, to generate a first synthetic version of the training subset of GSC, equal in number of audio files. We will refer, here-after, to the data as \textit{Synth} data in the result tables. It contains speech that were not properly synthesized (hallucinations and artefacts). In parallel, we generated another synthetic version, referred to as \textit{Synth (F)}, F standing for using the ASR-based filtering during the generation process, as explained also above. \textit{Synth (F)} data is expected to contain less poor speech material, if not none, and, thus, is expected to lead to better results.




\begin{table}[tbh]
    \centering
    \caption{Results obtained on GSC test set from training a set of MatchboxNet models with real and synthetic data. Each system was trained 5 times using different random seeds, the mean and standard deviation values are presented. Literature results are also reported for reference, 
    but they are not directly comparable since they use an additional silence class.}
    \vspace{0.25cm}
    \label{tab:results_matchboxnet}
    \begin{tabular}{cccc}
        \hline
        \textbf{System} & \textbf{Params.} &\textbf{Data}  & \textbf{Acc. (\%)} \\
        \hline
        \thead{BC-ResNet-8 \cite{bcresnet8}} & \thead{321k} & \thead{Real} & \thead{98.6} \\
        \thead{Emb. + head \cite{google_emb}} & \multirow{2}{*}{\thead{385k}} & \thead{Real} &  \thead{97.7}\\
        \thead{Emb. + head \cite{google_emb}} &  & \thead{Synth.} &  \thead{92.6}\\

        \hline
        \hline
        \multirow{3}{*}{\thead{MatchboxNet\\$3 \times 1\times 64$}} & \multirow{4}{*}{\thead{73k}} & \thead{Real} & \thead{$98.37\pm0.08$}  \\
        &   & \thead{Synth.} & \thead{$89.29\pm0.49$} \\
        &   &  \thead{Synth. (F)} & \thead{$92.06\pm0.35$} \\
        \hline
        
        \multirow{3}{*}{\thead{MatchboxNet\\$6\times 2\times 64$}} & \multirow{4}{*}{\thead{134k}} & \thead{Real} & \thead{$98.49\pm0.11$} \\
        &  & \thead{Synth.}  & \thead{$90.02\pm0.43$}\\
        &  &  \thead{Synth. (F)} & \textbf{\thead{$92.57\pm0.58$}}\\
        
        \hline
    \end{tabular}
\end{table}

\section{First SCC experiments}
\label{kws_experiments}

\subsection{Baseline SCC model: MatchboxNet}
\label{matchboxnet}

A MatchboxNet model \cite{matchboxnet} is a convolutional lightweight architecture specifically targeted for keyword spotting, command classification and wake-word detection. The architecture can easily be scaled in order to accommodate larger amounts of parameters under the form $B\times R \times C$ , where B stands for main quantity of convolutional blocks, R stands for sub-blocks (time-channel separable convolutional sub-blocks) and C for the amount of channels inside each block. In the present study, two different configurations of Matchboxnet models were used: $3 \times 1\times 64$ and $6\times 2\times 64$. These models can easily be adapted to a variety of domains and present state-of-the-art results on the GSC test set.

\subsection{Fully Synthetic SCC}
\label{kws syn}

Matchboxnet models were trained on three different subsets: either the original speech commands corpus (\textit{Real}), or the synthetic version (\textit{Synth}), or the synthetic version generated with ASR-based filtering (\textit{Synth. (F)}). A total of 50 training epochs with early stopping, a batch size of 128, a dropout of 0.25 and a cosine annealing learning scheduler, starting at 5e-3 and finishing at 5e-12, were used as main training hyperparameters. At each epoch, our models were validated using the real GSC validation set. Furthermore, all our models were similarly tested on the original test set. Accuracy values are reported in Table \ref{tab:results_matchboxnet}. As expected, the best results were obtained with the original GSC training data (\textit{Real}), with accuracy values above 98\%. Results obtained in the \textit{Synth.} condition, using synthetic data for training, achieved accuracy values of 89-90\%. The ASR-based filtering technique led to an accuracy gain of over two percentage points on MatchboxNet models, demonstrating that uncontrolled hallucinations in synthetic training data can contaminate the dataset and degrade model performance. 
A significant gap still remains between training high-performing models with real, original, speech data and purely synthetic data, generated with XTTS v2 and voice cloning.
We may compare these results to other approaches in the literature using synthetic data, namely \cite{google_emb}, with the limitation that they used the additional silence class, while we did not. Nevertheless, while the accuracy results remain quite similar (~92.6\%), our largest Matchboxnet model (134 k) is roughly one third smaller than their model sizes, in number of parameters.  


\begin{figure*}[!htb]
\centering
\begin{subfigure}{0.49\textwidth}
  \centering
  \includegraphics[width=\linewidth]{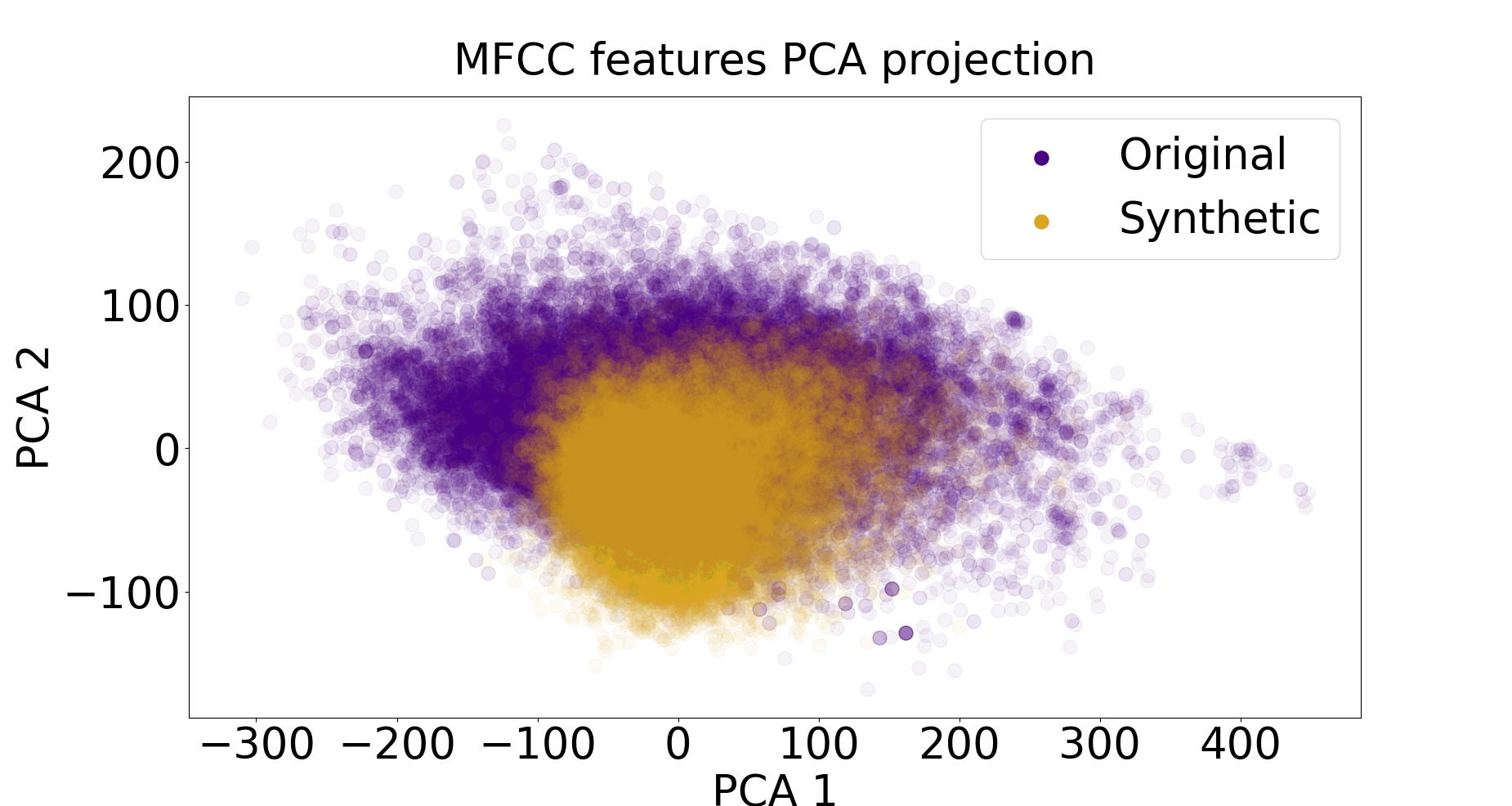}
  \label{fig:sub1}
\end{subfigure}
\hfill
\begin{subfigure}{0.49\textwidth}
  \centering
  \includegraphics[width=\linewidth]{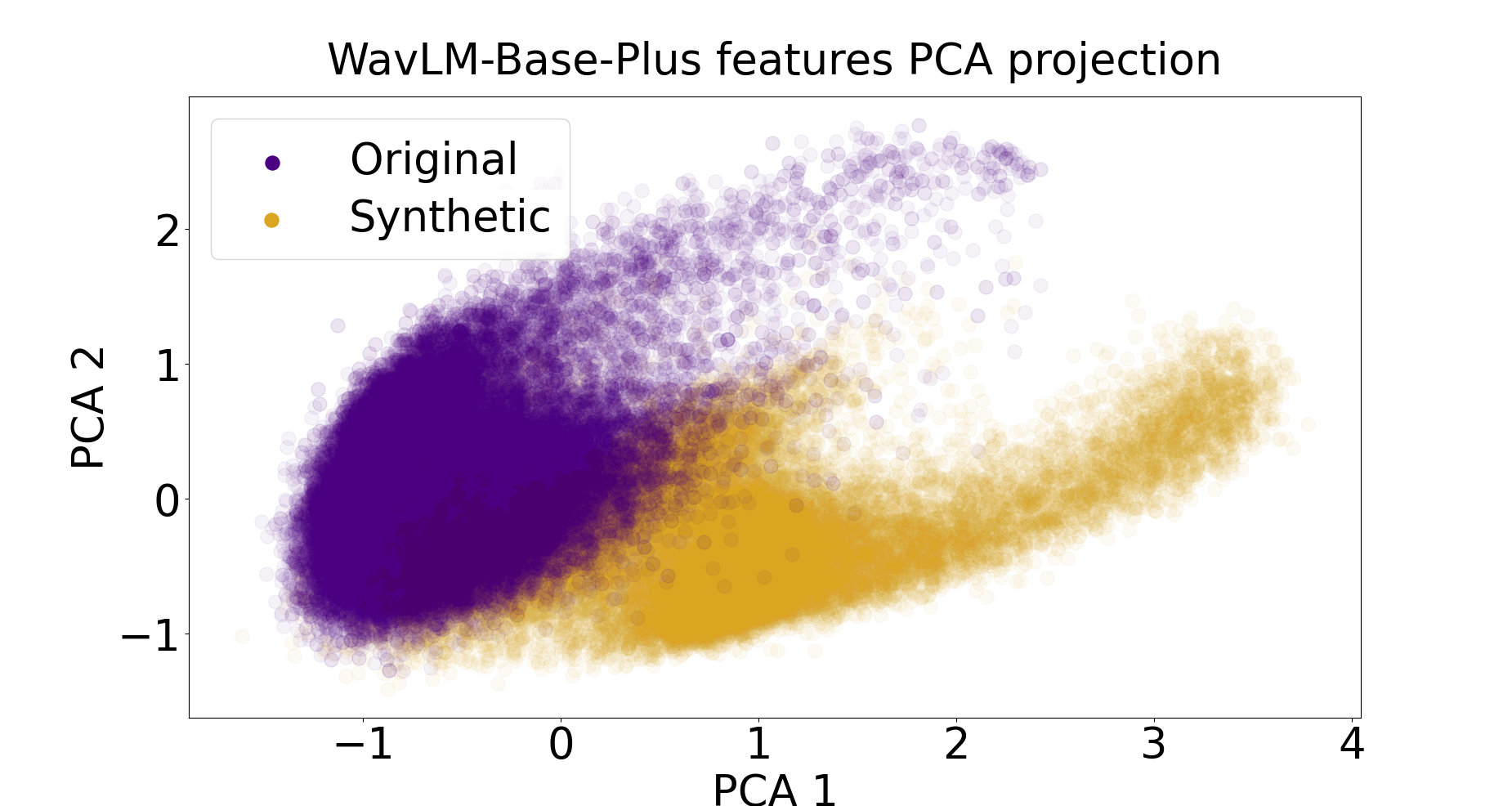}
  \label{fig:sub2}
\end{subfigure}
\caption{PCA analysis of filtered synthetic data and real speech data using Mel Frequency Cesptral Coefficients (MFCC) features (left) and WavLM self-supervised feature (right).}
\label{fig:pca}
\end{figure*}

\section{SSL-based SCC Experiments}

Despite the relatively high base accuracy of 92.57\% witnessed with our approach, there is still a gap of roughly 6\% between systems trained on real \textit{vs.} synthetic data. As an attempt to reduce this gap, we explored models based on SSL speech representations, trained with our synthetic data. Indeed, SSL benchmarking results already displayed strong results, in particular on GSC (v1)~\cite{benchmark}.



\subsection{Results with WavLM-Base-Plus representations}
\label{ssl_kws}


We used WavLM-Base-Plus to extract 768-d speech feature representations, from its twelfth transformer layer~\cite{wavlm}. Each audio excerpt is, in fact, represented by a sequence of such features, but we reduced the sequences into single 768-d vectors using statistic pooling, a method that computes the mean and standard deviation by feature dimension~\cite{stats_pooling}. A single linear layer is used as classification head for our task.

We trained the classification layer (WavLM-Base-Plus was frozen) over 30 epochs with a fixed learning rate of 5e-3 and a batch size of 128. As we did with the MatchboxNet models, we compare the three scenarios, using either the original GSC speech training data, or synthetic speech, without and with ASR-based filtering. The results, reported in table \ref{tab:results_ssl}, show that the performance gap between training a model either on real data or on the filtered synthetic speech has been reduced, compared to the previous experiments using MatchboxNet and MFCCs: 96.11\% (\textit{Synth.(F)}) \textit{vs.} 98.03\% (\textit{Real}). The \textit{Synth.(F)} SSL-based linear classifier is also much better that its  MatchboxNet counterpart, with an absolute gain of 3.5\% in accuracy. Interestingly, the results obtained when using the synthetic data without filtering (\textit{Synth.}) displayed a subpar accuracy of around 83\%, showing that hallucinated speech has a negative impact stronger when using SSL features compared to MFCCs.


\subsection{PCA Visualizations}

In order to get some insights between GSC real speech and our synthesized speech material, we visualize, in Fig.~\ref{fig:pca}), a 2-d dimension reduction through PCA over two sets of features: i) the 64 Mel-Frequency Cepstral Coefficients (MFCCs), as used by the MatchboxNet models, ii) the 768-d WavLM-Base-Plus feature vector representations, \cite{wavlm}. In both cases, statistic pooling over the time dimension was used to reduce sequences into single vectors. The MFCC plot (left plot) shows two superposed clusters, hinting that real speech is less clustered and more scattered than real speech. Surprisingly, with the SSL features, the two data points are much less superposed (right plot). They seem to even be linearly separable, showing that SSL features, even reduced to two dimensions, could be reliably used to differentiate between real and synthetic speech material. Given the impressive results witnessed when using SSL features on different tasks such as speech classification and KWS \cite{benchmark}, this clear separation in the two dimensional PCA space comes as unexpected. We draw the hypothesis that domain adaptation between the two types of speech material, in the SSL latent space, could help reduce the remaining performance gap  in our experiments.



\subsection{CycleGAN Domain Adaptation}

CycleGAN models~\cite{cyclegan} have been widely used to perform domain adaptation, in particular due to the increased training stability, compared to simpler GANs. They are a generative approach to convert inputs from one domain to another, and then convert them back to the original domain in a cyclic fashion. Our objective is to make closer in distribution the two domains that are the synthetic and real speech representations, in the 768-d SSL space. 

\begin{figure}[!t]
\centering
\includegraphics[width=3.3in]{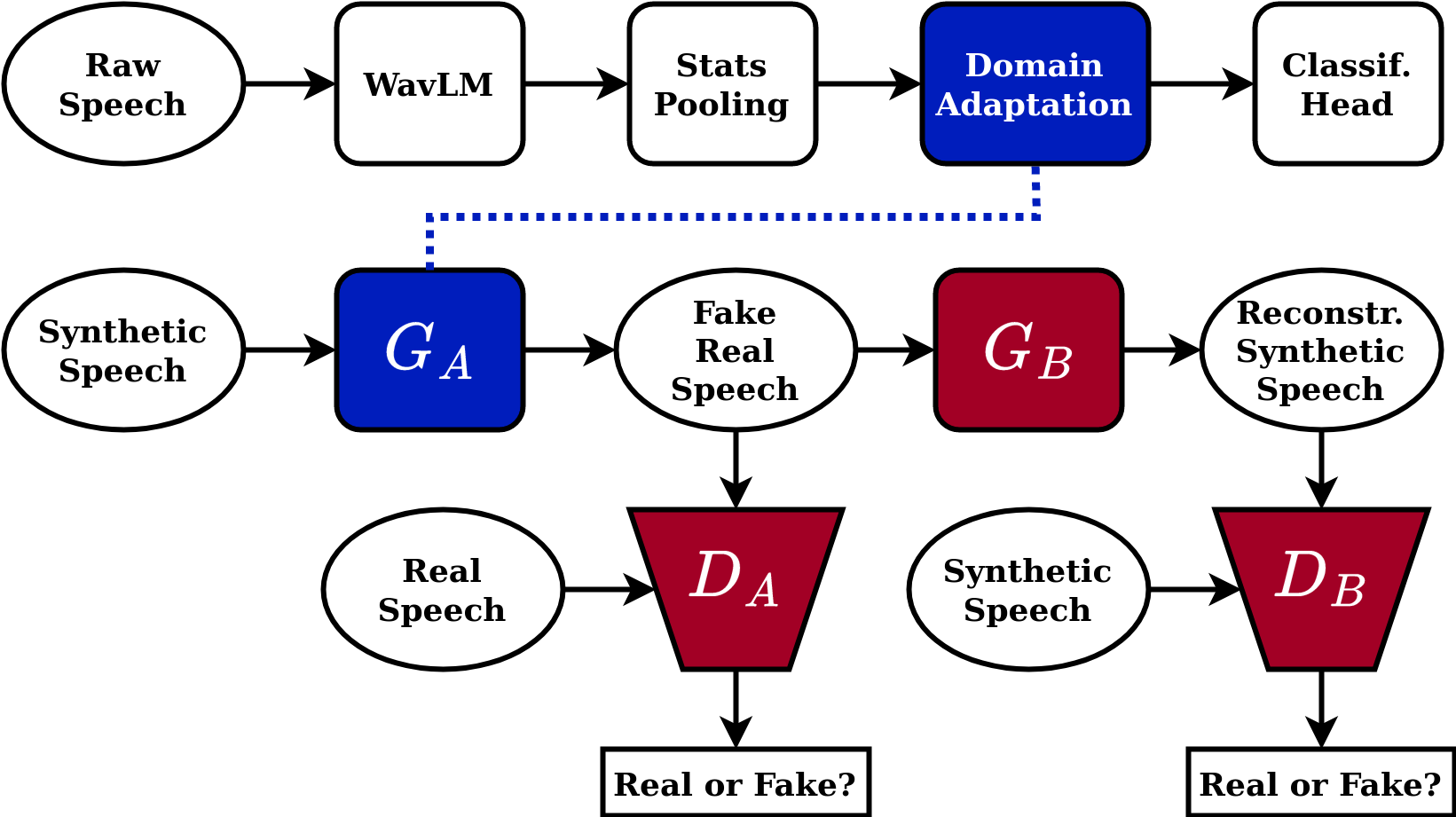}    
\caption{Proposed CycleGAN architecture and domain adaptation for a WavLM based SCC system.}
\vspace{-0.3cm}
\label{cyclegan}
\end{figure}

Our proposed CycleGAN, illustrated in Fig. \ref{cyclegan}, contains two generators ($G_A, G_B$) and two discriminators ($D_A, D_B$). The first generator aims to convert inputs from a synthetic domain to a real domain, and the second one the opposite. The two discriminators aim to differentiate between either real and generated real ($D_A$) or synthetic and reconstructed synthetic ($D_B$). Each generator contains three linear layers of dimensions [1536x512], [512x512] and [512x1536] respectively, with an intermediary ReLU non-linearity and a final hyperbolic tangent as activation function.  The discriminators share a similar architecture, with the only difference being in the final linear layer, adjusted to have dimensions of [512x1], and a sigmoid activation function.


The CycleGAN is optimized using a cycle-consistency loss function $\mathcal{L}_{CC}$ \cite{cyclegan} (see Eq. \ref{eq1}) that combines the discriminator and generator loss functions (MSE loss function, see equation \ref{eq2}) and a cyclic and identity loss with their respective weights. The cycle loss calculates the full reconstruction loss of the same input, enforcing consistency (eq. \ref{eq3}). Contrarily, the identity loss encourages the generator to produce outputs that are close to the inputs when the inputs are already in the target domain (Eq. \ref{eq4}). 

\vspace{-0.35cm}

\begin{equation}
\label{eq1}
\bm{\mathcal{L}_{CC}} = \mathcal{L}_{gan}^A+\mathcal{L}_{gan}^B+\lambda_{c}(\mathcal{L}_{c}^A+\mathcal{L}_{c}^B)+\lambda_{id}(\mathcal{L}_{id}^A+\mathcal{L}_{id}^B) 
\end{equation}




\vspace{-0.4cm}
\begin{equation}
\label{eq2}
\bm{gan:} A \xrightarrow{D_{A}(G_{A})} B  \;\;\;\;\;\;\;\;\; \textcolor{gray}{\text{[MSE Loss]}} 
\end{equation}

\vspace{-0.4cm}
\begin{equation}
\label{eq3}
\bm{cyc:} A \xrightarrow{G_{A}} B \xrightarrow{G_{B}} A,B  \:\;\;\; \textcolor{gray}{\text{[L1 Loss]}}
\end{equation}

\vspace{-0.4cm}
\begin{equation}
\label{eq4}
\bm{id:} A \xrightarrow{G_{B}} A \;\;\;\;\;\;\;\;\;\;\;\;\;\;\;\;\;\;\;\;\;\;\;\;\;\textcolor{gray}{\text{[L1 Loss]}}   
\end{equation}


Our CycleGAN is trained over 200 epochs using two data batches, one comprised of synthetic data and one of real data, using the 25 words of GSC classified as \textit{unknown} (see \ref{speech_commands}). The real and synthetic mini-batches are drawn randomly and independently. The system is optimized with a learning rate of 1e-5, a batch size of 128, a lambda cycle ($\lambda_{cyc}$) of 10.0 and a lambda identity ($\lambda_{id}$) value of 0.5. Since our main goal is to make synthetic speech representations closer to real ones, only the generator A is used to perform the domain transformation, as illustrated in Fig. \ref{cyclegan}.

Similarly to section \ref{ssl_kws}, linear classifiers are trained using the same hyperparameters as before, this time using the pre-trained domain adaptation provided by the CycleGAN, with a frozen Generator A. The results, reported in Table \ref{tab:results_ssl}, show a small but noticeable improvement in accuracy when using our CycleGAN versus the system trained without this domain adaptation.

\begin{table}[!htb]
    \centering
    \caption{Results obtained on the GSC test set from training a set of WavLM-based linear classifiers using real data, synthetic data (without and with filtering) and synthetic data transformed with our CycleGAN. Each system was trained 5 times using different random seeds, the mean and standard deviation values are presented}
    \vspace{0.25cm}
    \label{tab:results_ssl}
    \begin{tabular}{ccc}
        \hline
        \textbf{System} & \textbf{Data}  & \textbf{Acc. (\%)} \\
        
        \hline
        
        \multirow{4}{*}{\thead{WavLM Base Plus \\w/ stats pooling}} &  \thead{Real} &  \thead{$98.03\pm0.11$}\\ 
         &   \thead{Synth.} &  \thead{$83.05\pm0.82$}\\
         &   \thead{Synth. (F)} &  \thead{$96.11\pm0.08$}\\   
         &   \thead{Synth. (F) + GAN} & \thead{$96.51\pm0.11$} \\  

        \hline
    \end{tabular}
\end{table}

\section{Discussion}
\label{discussion}

Despite advancements in the quality of modern TTS systems, PCA analysis on MFCCs --- and even more so on SSL features (Fig. \ref{fig:pca}) --- revealed noticeable differences in cluster size and position between synthetic and real speech data. This indicates that synthetic speech still lacks the variety and diversity inherent in real speech, but also a mismatch between the feature representation spaces, at least with SSL features. While the literature and our experiments show good performance when using SSL speech representations for our task, it was surprising to observe such a distinct separation between synthetic and real speech data. Despite the CycleGAN methodology promoting a small performance gain, a number of questions appear, left for future work. We aim to conduct an in-depth analysis of the SSL features to identify which specific features are most relevant in distinguishing between synthetic and real speech data. We could potentially exclude them in our downstream application, thereby aligning it more closely with real data characteristics.  
A deeper analysis of the domain adaptation actually achieved by the CycleGAN is essential to evaluate how closely the transformed data distribution aligns with the real data. The modest performance gain obtained with CycleGAN adaptation suggests that the distributions were brought closer together. However, we would like to further quantify this alignment more rigorously. Additionally, exploring other recent domain adaptation methods, such as Flow Matching~\cite{lipman2022flow}, could provide further insights and potentially enhance the adaptation process.

We are also interested in exploring the impact of potential distribution differences between synthetic and real speech on other downstream tasks, such as emotion recognition and intention detection. In these tasks, fixed-length representations could be adapted from one domain to another (\textit{e.g.}, converting neutral speech to various emotional states), especially for systems based on self-supervised learning features. This approach could lead to improved performance and the development of new data augmentation strategies.

Additionally, our current study is limited to a single TTS system that, although widely used and prominent in the speech community. Testing other recent TTS systems, such as WhisperSpeech\footnote{https://github.com/collabora/WhisperSpeech}, would be a valuable future extension to this work.






\section{Conclusion}
\label{conclusions}
 
We used XTTS v2, a recent end-to-end TTS system with voice cloning capabilities, to generate synthetic speech commands data that mimic the training subset of the standard Google Speech Commands dataset. We sourced speech excerpts from Common Voice multilingual datasets, ensuring a large enough sample to assign a unique speaker to each generated audio clip through voice cloning.  

The lightweight MatchboxNet model (134k parameters), trained on this synthetic data, achieved 90.0\% accuracy on the test set, which improved to 92.6\% when an ASR-based filtering technique was employed to remove poorly synthesized audio samples. These results are comparable to other methods in the literature but use significantly fewer parameters. 
Nevertheless, this result remains below the 98.5\% accuracy achieved using the real training set. Using the synthetic data (with filtering), we achieved a higher accuracy of 96.1\% by leveraging self-supervised representations from WavLM-Base-Plus, which were vectorized with a statistical pooling layer and classified using a fully connected layer.

A PCA plot of the WavLM SSL features showed that the distributions of the synthetic and real speech feature representations do not fully align. We proposed using a CycleGAN for domain adaptation to transform synthetic speech representations into more realistic ones. This approach further improved the accuracy to 96.5\%, narrowing the performance gap between models trained on synthetic and real speech data. 



\vfill\pagebreak

\bibliographystyle{IEEEbib}
\bibliography{strings,refs}

\end{document}